# Seeing Quantum Fractals

Gregory A. Fiete and Alex de Lozanne
Department of Physics, University of Texas at Austin, Austin TX 78712
USA

Anyone fortunate enough to have been on a holiday flight that lands near a beautiful sunny coastline will no doubt have taken in the scenery on the way down, and perhaps even marveled at the winding of the shore into the horizon. What you may not have realized is that the coastline is actually a fractal—an object that appears the same on all length scales (perhaps only statistically). Fractals actually abound in nature: Galaxies, clouds, mountains, trees, and broccoli are all familiar examples. But fractals can occur in the quantum realm as well, even though they have never been observed there, until perhaps now. On page 665 of this issue of Science, Richardella et al report direct measurements of quantum mechanical electron waves that indicate they may possess a fractal nature.

Quantum particles, such as the electrons surrounding the nuclei of atoms, have a wave nature associated with them that gives rise to interference phenomena much like that found in waves on the surface of water, light waves, or sound waves. Ordinarily these quantum mechanical electron waves are extremely difficult to see because their ripples occur on the scale of the spacing of individual atoms in a solid. However, a powerful experimental tool known as the scanning tunneling microscope (STM) has the capability to locally image these quantum waves on the surface of a material (1,2). As the STM is able to measure these atomic-scale waves over a fairly wide range of energies and with good energy resolution, it is an ideal tool to crack probe quantum mechanical effects in solid materials.

The crucial physical ingredient that gives rise to fractals in electron waves is the phenomenon of localization of waves in a disordered medium (3). A light wave traveling in vacuum will continue traveling indefinitely. However, waves traveling in a random environment can get "stuck" or "localized" in certain positions due the complicated pattern of constructive and destructive interference taking place from random scatterings. It turns out that in many common situations a weakly random background will still allow wave propagation. However, when the fluctuations in the random environment become too large the waves localize, *i.e.* they are no longer spread over the entire system but only occupy a small region. There is thus a special value of the randomness where the waves "decide" to make the switch from extended over the system to localized. At this transition point, the waves are fractal (4,5). In the context of electron waves in a solid, the transition from extended waves to localized waves as a function of disorder is called the metal-insulator transition, and its physics is the focus the STM study reported by Richardella et al. (see figure in published version).

The material chosen for this work was $Ga_{1-x}Mn_xAs$, a ferromagnetic semiconductor intensely studied over the last decade because of its potential use in electronic devices based on the quantum mechanical "spin" of the electron. While the non-magnetic semiconductor GaAs has been an industry standard for decades, the field of ferromagnetic semiconductors opened up when it was shown that doping the large-spin transition metal Mn into GaAs could result in a ferromagnetic material (6) that could readily be interfaced with GaAs to allow the efficient injection of spin, a requirement for spin-based devices (7, 8). The Mn introduced into GaAs disorders it, and at the same time supplies free carriers that form the quantum waves undergoing a metal-insulator transition as a function of Mn concentration, x (9). While the magnetic properties of $Ga_{1-x}Mn_xAs$ were not the focus the work of Richardella et al, they are intimately related to the electrical properties because the carriers mediate the magnetic interaction and therefore play a direct role in the magnetism itself. Obtaining a detailed understanding of the electronic states and their precise relationship to the ferromagnetic properties of $Ga_{1-x}Mn_xAs$ is the central challenge in this field. The metal-insulator transition presents an extra degree of difficulty in this quest due to the inherent large fluctuations, but also grants an opportunity for intriguing quantum phenomena to emerge.

The main contribution of the work of Richardella et al is to provide an energy-resolved real-space measurement of quantum wave properties near the metal insulator transition in $Ga_{1-x}Mn_xAs$. While some features of their data are suggestive of fractal aspects of the quantum waves, a number of questions are raised by their experiments. For example, why does the carrier concentration appear to be nearly independent of the Mn concentration? What is the physics that leads to this preferred carrier concentration? Why do the strongest signatures of the metal insulator transition appear at one particular energy, independent of Mn concentration? How much do these surface measurements reveal about the bulk physics? Taken as a whole, the data indicate that interactions among the carriers play an important role in this system. Complementary studies of the metal-insulator transition in electrical transport have reached a similar conclusion (10). It thus appears likely that interaction effects at the metal-insulator transition are here to stay, and may even lead to novel effects like "self-organized quantum criticality" (11) on the surface of a recently discovered class of materials known as "topological insulators" (12,13). For the time being, it looks like we might have to be content with a clear view of fractals from 35,000 feet up.

**Research article** is A. Richardella et al. Science 327, 665 (2010).